\documentclass[showpacs,floatfix,preprint,nofootinbib]{revtex4}

\begin{document}

\title{X rays test the Pauli exclusion principle}
\author{A.~Yu.~Ignatiev}
\address{ 
    \em School of Physics, Research Centre for High Energy Physics,
     University of Melbourne,   Australia}
\email{a.ignatiev@physics.unimelb.edu.au}

\pacs{03.65.-w, 11.30.-j, 31.90.+s}

\def\be{\begin{equation}}
\def\ee{\end{equation}}
\def\bea{\begin{eqnarray}}
\def\eea{\end{eqnarray}}
\newcommand{\nn}{\nonumber \\}

\begin{abstract}
Since the publication of the models describing a small violation of the 
Pauli exclusion principle (PEP) there has been an explosion of word-wide 
interest in PEP tests and related theories. 

PEP forbids an atom to have more than 2 electrons in the K-shell. If PEP 
is slightly violated, a third electron can occasionally join in. This 
would result in an anomalous X-ray emission. A high-sensitivity 
experiment places an upper limit of the order of $10^{-26}$ on the PEP 
violating parameter. 

I will outline the main theoretical and experimental ideas in this new 
exciting area. 

\end{abstract} \maketitle

\section{Introduction}
 The Pauli exclusion principle (PEP) is one of the pillars of physics and 
 chemistry
and we know from history that pillars can crack. The most familiar 
examples are parity and CP violation. For this reason alone we should be 
willing to look into possible PEP violations. But what are other, more 
specific reasons for looking into it? The main reason is that our 
theoretical understanding of the PEP origin is still not quite 
satisfactory.  Perhaps the best way to illustrate it is to quote Feynman 
(1965): ``Why is it that particles with half-integral spin are Fermi 
particles whereas particles with integral spin are Bose particles? We 
apologize for the fact that we cannot give you an elementary explanation. 
This probably means that we do not have a complete understanding of the 
fundamental principle involved.'' Similar dissatisfactions were expressed 
by Pauli (1955) and Dirac (1981).

 One way to look 
for more understanding is to try to think what happens if the principle 
is violated. The 3 main questions to be considered (of course, they are 
interrelated) are: 

Can we make up a theory that would deviate from PEP by a small amount?

How to look  for possible PEP violations experimentally? 

What number characterises the accuracy with which PEP holds?

The last question in fact shows how different PEP is from other similar 
principles, for example CPT symmetry. For CPT symmetry there is a well 
defined parameter that describes possible deviations from CPT such as the 
mass difference between a particle and antiparticle. As for PEP, we had 
not had such a parameter until we started to make up theories that would 
be able to describe its violation and the theories then showed us how to 
introduce such a parameter. The purpose of this talk is to give an idea 
of what the answers to these 3 questions we have at the moment. 

What are other differences between PEP violation and the violation of 
more familiar symmetries such as CP? In the case of CP violation we can 
write a lagrangian which is CP violating and this is a pretty 
straightforward procedure in QFT. In other cases we can write a 
lagrangian that is symmetric under some operation and then add a small 
part to break down the symmetry. But in the case of PEP violation we 
cannot write a PEP violating lagrangian. PEP can be violated only through 
the commutation relations and not through the lagrangian. So this was 
really quite a new situation to consider  and for that reason it was 
rather challenging: ``you can't be a little bit pregnant". 

There is one question that arises almost always in discussions of PEP 
violation: what about the spin-statistics theorem? Doesn't this theorem 
rule out the possibility of PEP violation straight away? The answer to 
this question is no.  The spin-statistics theorem of the axiomatic 
quantum field theory forbids quantising the fields with half-integer spin 
according to Bose, that is with commutators.  This theorem leaves open 
the question of possible alternative ways of quantising the spin one-half 
fields, in particular it says nothing about  those new ways which could 
lead to small PEP violations.  A recent review of attempts to derive PEP 
from other principles of quantum mechanics is given by Kaplan (2002).  

No theoretical models of small PEP violation existed up until 1987. 
Although non-standard types of statistics, such as parastatistics (Green, 
1953), had been known, they could be viewed as ``100\%  violation'' of 
PEP rather than small violation. This kind of violation for electrons and 
nucleons  was clearly ruled out by experiment. Also, there were studies 
of possible small violation of electron identity (Luboshitz and 
Podgoretskii, 1971; Okun (1989); see also Fischbach et al., 1968 and 
Fermi, 1933). Although related to small PEP violation, small non-identity 
is a different idea. 

This is a brief outline of a large and growing area. In particular, the 
ten-page limit does not allow for complete referencing.  More detailed 
reviews and further references can be found in the Proceedings of the 
Conference ``Spin-statistics connection and commutation relations'' 
(Hilborn and Tino, 2000). To the extent allowed by the material, the 
present review is complementary rather than overlapping with the previous 
ones. An extensive bibliography with hundreds of references has been 
compiled by  Gillaspy and Hilborn (2000).  

\section{Theories}

Let us consider several formulations of the Pauli exclusion principle 
which we use today and which are rather different from the original 
formulation given by Pauli in 1925: ``There can never be two or more 
equivalent electrons in an atom. These are defined to be electrons for 
which $\dots$ the values of all quantum numbers $\dots$ are the same.'' 

The second formulation came with the advent of Quantum Mechanics in our 
modern version. This formulation is due to Dirac and it states that the 
wave function of the system of electrons must be antisymmetric under 
exchanges and permutations of electrons. In the simplest case we have to 
require that the wave function of two electrons is antisymmetric with 
respect to the exchange of their space and spin coordinates. 

 Finally, the 
most general formulation of PEP is given by the second quantisation 
formalism which is used in quantum field theory to describe electrons and 
positrons that can be created and annihilated. The formulation in terms 
of second quantisation concepts requires that the fields describing the 
electrons, positrons or any other particles with half-integer spin obey 
anticommutation relations of the form $AB+BA=0$ or another c-number. The 
main thing  is that we have here plus sign rather than minus sign when we 
interchange $A$ and $B$ operators. The minus sign would correspond to 
Bose commutation relations. 

Logically, the third formulation is the most general and it implies the 
second formulation which in turn implies the first one. Because it is the 
most general, it is the best formulation to work with although it sounds 
a bit more abstract that the other two. 

In summary, the validity of the Pauli principle follows automatically 
from the form of the anticommutation relations between the operators of 
the electron-positron fields $\psi(x)$, $\bar{\psi}(x)$ or, which is 
equivalent, between the creation and annihilation operators of the 
electrons and positrons $a_{k\sigma}, a_{k\sigma}^{\dagger}, b_{k\sigma}, 
b_{k\sigma}^{\dagger}$. 

Therefore in a theory, describing the violation of the principle the 
anticommutation relations should certainly be  changed.  Apriori it is 
not  clear at all in what form  one should cast new commutation 
relations.  

Generally speaking, to solve this problem one could reason as follows: 
consider various sets of the commutation relations between the creation 
and annihilation operators $a_{k\sigma}, a_{k\sigma}^{\dagger}, 
b_{k\sigma}, b_{k\sigma}^{\dagger}$ or  fields $\psi(x), \bar{\psi}(x)$ 
of the most general form. One should consider not only bilinear 
relations, but also trilinear ones and so on, each set containing not 
necessarily one but, in general, several independent relations.  

Furthermore, one should require that these relations satisfy a number of 
general principles of the quantum field theory:  positivity of energy, 
Lorentz invariance, relativistic causality, the electric charge 
(fermionic number) conservation, and  C-parity.  As for the last four 
principles, one should not require that these principles be absolutely 
valid, it suffices to require that their possible violations were at most 
of order O($\beta$) (where $\beta$ is the Pauli principle violating 
parameter) so that when $\beta \rightarrow 0$ all these violations were 
unobservably small. 

In this way one could in principle find all the allowed commutation
relations (or prove that they do not exist) containing a small
parameter $\beta$ and resulting in the usual fermi-statistics of the
electrons when $\beta \rightarrow 0$. 

In practice, however, this way seems intractable.  A simpler method 
consists in the following (Ignatiev and Kuzmin, 1987). Instead of 
searching for the algebra of the operators, let us try to construct first 
the representation of this algebra possessing the necessary property and 
after that try to find the commutation relations themselves.  It is more 
convenient to work with the creation and annihilation operators than with 
the field operators.  Let us further simplify the problem by discarding 
the momentum and spin variables and considering only electron (but not 
positron) operators. 

Thus, we should find representation of the creation and annihilation
operators $a$, $a^{\dagger}$ depending on the parameter $\beta$ so that 
when $\beta$ tends to zero this representation goes over  into the usual 
Fermi representation 
\begin{equation}
\label{eq:2}
a_F = \left( \begin{array}{ll}
0 & 1 \\
0 & 0 
\end{array} \right), \qquad
a_F^{\dagger} = \left( \begin{array}{ll}
0 & 0 \\
1 & 0 
\end{array} \right).
\end{equation}

As the orthonormal basis here we take the vacuum $|0\rangle$ and the
one-particle state $|1\rangle$.  Evidently, the minimal dimensionality of 
the state space we are looking for is three, so we choose as the basis of 
that space the states $|0\rangle$ (vacuum), $|1\rangle$ (one-particle 
state) and $|2\rangle$ (two-particle state).  Suppose that the action of 
the creation and annihilation operators is defined as follows (the 
parameter $\beta$ is supposed to be real): 
\begin{eqnarray}
\label{eq:3}
 a^{\dagger}|0\rangle = |1\rangle &\qquad& a|0\rangle = 0\nonumber\\
 a^{\dagger}|1\rangle = \beta |2\rangle &\qquad& a|1\rangle =
 |0\rangle\\
 a^{\dagger}|2\rangle = 0 &\qquad& a|2\rangle = \beta|1\rangle. \nonumber
\end{eqnarray}

Then the matrices of these operators in the chosen basis take the 
following form: 
\begin{equation}
\label{eq:4}
 a= \left( \begin{array}{lll}
0 & 1 & 0 \\
0 & 0 & \beta \\
0 & 0 & 0
\end{array} \right) \qquad
a^{\dagger} =  \left( \begin{array}{lll}
0 & 0 & 0 \\
1 & 0 & 0 \\
0 & \beta & 0
\end{array} \right). 
\end{equation}

The Hilbert state space $H$ can be decomposed into the direct sum of the 
subspaces $H_2$ (built on the vectors $|0\rangle, |1\rangle$) and $H_1$ 
(built on the vector $|2\rangle$).  It is clear that if $\beta=0$ the 
transitions between the states in $H_2$ and $H_1$ become forbidden so 
that the space $H_1$ gets completely decoupled from $H_2$. 

Now, let us construct the commutation relations (i.e., the algebra) which 
is satisfied by the operators $a$, $a^{\dagger}$.  To do that, one should 
calculate various products of the operators $a$, $a^{\dagger}$ of the 
form $a^2$, $a^{\dagger}a$, $a a^{\dagger}$, $a^3$ etc. and then find the 
relations between them (such relations should certainly exist because 
there are only 9 independent 3$\times$3 matrices).  

First, it can be shown  
that the operators $a$, $a^{\dagger}$ and their bilinear products are linearly independent, i.e. the 
bilinear commutation
relations are absent in the model under consideration.

Next, the trilinear relations do exist and can be
written, for example, in the following form:
\begin{eqnarray}
a^2a^{\dagger} + \beta^2 a^{\dagger}a^2 &=& \beta^2 a\label{eq:9}\\
a^2a^{\dagger} + \beta^4a^{\dagger}a^2 &=& 
\beta^2aa^{\dagger}a,\label{eq:10} 
\end{eqnarray}
plus their Hermitean conjugate relations.  To these relations one should 
add the equalities 
\begin{equation}
\label{eq:11}
a^3 = 0, \qquad (a^{\dagger})^3 = 0.
\end{equation}
Thus the equalities (\ref{eq:9}) - (\ref{eq:11}) complete the 
construction of our algebra. 

Next let us find the particle number operator $N$ in this model. In the 
chosen representation the operator $N$ has the form 
\begin{equation}
\label{eq:12}
N = \left( \begin{array}{lll}
0 & 0 & 0 \\
0 & 1 & 0 \\
0 & 0 & 2
\end{array} \right)
\end{equation}
so that the usual commutation relations hold true
\begin{equation}
\label{eq:13} [N,a] = -a, \quad [N,a^{\dagger}]=a^{\dagger}. 
\end{equation}

Is it possible to find the bilinear expression for the operator $N$ in 
terms of the creation and annihilation operators? The positive answer is   
given by the form 
\begin{equation}
\label{eq:14} N=A_1 a^{\dagger}a + A_2 aa^{\dagger} + A_3  
\end{equation}
with the coefficients $A_i$ given by: 
\begin{eqnarray}
\label{eq:15} A_1 &=& \frac{-1+2\beta^2}{1-\beta^2+\beta^4} \nonumber\\ 
A_2 &=& \frac{-2+\beta^2}{1-\beta^2+\beta^4} \\ A_3 &=& 
\frac{2-\beta^2}{1-\beta^2+\beta^4}. \nonumber 
\end{eqnarray}

Thus we have completed the construction of the algebra of the creation 
and annihilation operators and also have found the bilinear expression 
for the particle number operator (Ignatiev and Kuzmin, 1987). 

Fundamental mathematical properties of the IK algebra have been studied 
by Biedenharn et al.(1989) (see also Ignatiev, 1990) and Cougo-Pinto 
(1993). These studies revealed interesting connections with such concepts 
as Jordan pairs, $C^*$-algebras, and quantum groups. 

      The problem of a 
realistic generalization of the one-level IK model can be formulated as 
the following question: how to write down the commutation relations if we 
ascribe the momentum and spin indices to the creation and annihilation 
operators? One way to do this was suggested by Greenberg and Mohapatra 
(1987, 1989a). The main problem is to make sure that no more than 2 
electrons can occupy the same state. In the 1-level model that was 
achieved by requiring $a^3=0$. However, in the multi-level model we have 
infinitely many possibilities to put 3 electrons in the same state 
because we can ``sandwich'' other electrons between them. These 
sandwiched states turn out to have negative norms (Govorkov, 1989, 1983;
Greenberg and Mohapatra, 1989b). No way out of this difficulty has been 
found and it is believed to be a fatal flaw. In principle, the 
possibility of curing this theory cannot be completely ruled out. A 
well-known example is the ordinary QED which involves negative-norm 
states that are harmless. 

Another generalisation was attempted by Okun (1987) who assumed that the 
operators from different levels obey the usual anticommutation relations. 
This model also ran into serious difficulties discussed by the author.

 We would like to stress that the  ``small PEP violation'' is an 
 intuitive concept and one  can try to formalise it in many different ways leading 
 to different theories with their specific experimental predictions. 
 
 In particular, the theory called ``quons''  was proposed in (Greenberg, 
 1990, 1991; Mohapatra, 1990; Fivel, 1990, see also Biedenharn, 
 1989 and Macfarlane, 1989). This theory is based on {\em bilinear}  
 rather than trilinear commutation relations:
 \begin{equation}
 a_k a^{\dagger}_l - q a^{\dagger}_l a_k = \delta_{kl}.
 \end{equation}
 
 The main physical feature of the model is that there is no limit on the number 
 of particle that can occupy the same state, i.e., all types of Young tables are
 allowed  for a system of quons. A review of other properties of quons have been
 given in (Greenberg, 2000, see also Chow and Greenberg, 2001). 

Cosmological consequences of possible PEP violation for neutrinos have 
recently been discussed by Dolgov and Smirnov (2005); Dolgov et al. 
(2005); see also Cucurull et al.(1996). 
\section{Experiments}

The physics community, both theorists and experimentalists, took up the 
idea of small PEP violation with great enthusiasm. {\em Scientific 
American} published an article (Kinoshita, 1988) entitled ``Roll Over, 
Wolfgang?'' that described our work and the subsequent theoretical 
development by Greenberg and Mohapatra as well as the plans to look for 
small PEP violations experimentally. These experimental efforts were 
ongoing soon after the appearance of the theory papers. 

It is interesting to compare the number of papers published on the 
subject of small statistics violation before and after 1987. During 30 
years prior to 1987 there were about a dozen of papers on the topic 
including pioneering works by Reines and Sobel (1974), Logan and Ljubicic 
(1979), Amado and Primakoff (1980) and Kuzmin (1984) while since 1987 the 
number of papers has grown to several hundreds. 

 We are now coming to our main Question 2: How to look for small PEP 
violation in experiment? There are plenty of experiments of different 
kinds which have tried to look for deviations from PEP and it is hard to 
mention all of them, but most of them are based on one of two broad 
ideas. 

The first idea, originated by Reines and Sobel (1974) (see also Goldhaber 
and Scharff-Goldhaber, 1948), is to look for forbidden transitions to the 
levels occupied by particles such as electrons or nucleons. So we can 
look  for anomalous transitions to filled shells. As a result, the energy 
would be emitted in the form of X rays, gamma rays or something else. 

A more recent idea is to look for these anomalous states themselves 
(``the integral method''): for example, we can try  to look for atoms 
with 3 electrons in the K shell.  The first experiment of this kind was 
proposed in (Ignatiev and Kuzmin, 1988; Gavrin et al., 1988). The 
integral method was further developed by Novikov and Pomansky (1989), 
Novikov et al. (1990), and Nolte et al. (1991).  Similar ideas were 
discussed by Okun (1987, 1988). 

Let us sketch the first type of experiments. We have here 2 levels and 
one electron is on the upper level while another is on the lower level. 
Of course, the upper electron can go down to the lower level provided the 
proper selection rules are observed and to emit radiation, e.g. X rays 
when the lower level is the K shell of an atom. This is an allowed 
transition. 

Now, if there are 2 electrons with opposite spins on the lower level then 
all states on this level are occupied if this is a K shell. Then the 
upper level electron cannot go down to the lower level and this is a 
forbidden transition in the standard theory. 

Next let's take a look at the theory where PEP is slightly violated. Then 
this same transition to the fully occupied K shell can now be allowed and 
the emitted radiation can be a signature of this process so that we can 
try to look for X rays, for example, as  a sign of PEP violation. 

The concrete realisation of a type 1 experiment (which takes into account 
the Amado-Primakoff (1980) arguments) is the experiment performed  by 
Ramberg  and Snow (1990) at Maryland University. This is a remarkably 
simple, table-top kind of experiment. It has a thin copper plate through 
which a large electric current is sent and close to this strip of copper 
there is an X ray detector whose task is to detect the emission of X rays 
when the transition occurs to the K shell of a copper atom. The 
experiment shows that there are no anomalous X rays in this case and the 
upper limit can be deduced on the PEP violating parameter $\beta^2$.  
This limit turns out to be exceptionally low:  $\beta^2 \leq 1.7\times 
10^{-26}$. 

An improvement of this limit by 4 orders of magnitude is currently being 
planned by the VIP collaboration (2005) which builds on the success of 
DAFNE Exotic Atom Research program (DEAR) completed in 2003. The plan is 
to utilise the excellent X-ray detector involved in that program for the 
new purpose of the PEP violation search. The VIP setup will be first 
transported and installed at the Gran Sasso low-background laboratory ( 
LNGS); then (in 2005-2006) the data taking will proceed, alternating 
between periods of current on (``signal'') and off (``background'').

So far we were considering type 1 experiments that were looking for 
anomalous transitions accompanied by radiation. Now let's have a look at 
the alternative scheme where the search is for the anomalous states 
themselves such as, for example, anomalous elements. The first question 
that we have here is: what elements could make good candidates for such a 
search? 

One approach to answering is to consider chemical differences between the 
ordinary element with the atomic number $Z$ and the anomalous one with 
the same atomic number which we denote by a prime $Z'$. We know that the 
chemical properties of the elements are controlled by the number of 
valence electrons, i.e. the number of electrons in the outermost shell of 
the atom. 
 
 Now, the element $Z'$ has one valence electron less than  the ordinary element $Z$ because one 
elecrtron in $Z'$ goes down to the K shell. For light elements (roughly, 
$Z < 20$) the preceding element $Z-1$ in the periodic table also has one 
valence electron less than $Z$. From that we conclude that the chemical 
properties of the light elements with the anomalous K shell structure 
(i.e., having 3 electrons in it) are similar to the chemical properties 
of the ordinary element with the atomic number $Z-1$, the previuos 
element in the Mendeleev table. 

To make this idea more transparent, let's consider a specific example and 
let's take neon as our $Z$ element ($Z=10$), then our $Z-1$ element will 
be fluorine ($Z=9$).  Neon has $1s, 2s, 2p$ shells all completely filled 
and if one electron from $2p$ shell goes down to $1s$ shell violating the 
Pauli principle, then we'll have one electron less in the $2p$ shell.  
Let's now take a look at the electronic structure of the ordinary 
fluorine which precedes neon in the periodic table. The outer, i.e. $2p$ 
shell of fluorine has exactly the same structure as that of anomalous 
neon, $Ne'$. We can therefore conclude that chemically the anomalous neon 
would behave similarly to the ordinary fluorine. Symbolically, we can 
write it as $Ne'\sim F$. 
 
 Why do these elements make good candidates for experiments? Our task is to find element $Z$ so that 
there is a sharp contrast between the chemical properties of the elements 
with atomic numbers $Z$ and $Z-1$. One solution to this problem is that 
we take noble gases as our element $Z$, such as $Ne$ and $Ar$, and 
therefore $Z-1$ element will be a halogen: fluorine or chlorine.  The 
difference between these 2 groups of elements (noble gases and halogens) 
is great: the noble gases are very inert elements while  halogens are 
very active ones. So we have 2 solutions to our problem: $Z=10$, i.e. 
$Ne$--$F$ pair, with $Ne'$ behaving as $F$, and $Z=18$, $Ar$ -- $Cl$ 
pair, with $Ar'$ behaving as $Cl$.      

The experiments based on the idea that we have just discussed have been 
performed by Novikov et al. (1990) and Nolte et al. (1991) who looked for 
anomalous neon in a sample of fluorine and anomalous argon in a sample of 
chlorine. Schematically, their procedure is as follows: they first take a 
sample containing fluorine. Then they turn neutral fluorine atoms into 
negative ions. Finally, they use atomic mass spectroscopy to look for 
ions with atomic mass 20. 
 
 If these $A=20$ ions are found, one can  interpret it as anomalous $Ne$ atoms because of 2 reasons: 
first, fluorine does not have stable isotopes with A=20, so it cannot be 
a fluorine atom. Neon, on the other hand, cannot form negative ions, so 
these atoms cannot be neons either. Therefore, they are to be interpreted 
as anomalous neon atoms. This experiment obtained an upper limit on the 
concentration of $A=20$ and $A=36$ ions (Nolte et al., 1991):
\begin{equation}
\label{eq:nolte} N(^{20}Ne')/N(^{20}Ne) <2\times 
10^{-21},\;\;\;N(^{36}Ar')/N(^{36}Ar) <4\times 10^{-17}. \end{equation} 
 
 Another method was used in a search for anomalous helium, $He'$ 
(Deilamian et al., 1995). While ordinary He has a wave function that is 
antisymmetric with respect to space/spin exchange, the anomalous helium 
would have a {\em symmetric} wave function under this exchange. Because 
of this difference in symmetry the energy level of He' would be shifted 
by a tiny amount.  Drake (1989) calculated these tiny energy shifts with 
high precision. Therefore we know the exact position of an anomalous 
spectral line. The idea of the experiment is to look for the anomalous 
spectral  line ($1s3p ^{1}P_1 \rightarrow 1s2s ^{1}S_0$) using the modern 
Doppler-free laser spectroscopy technique. The absence of this spectral 
line allows one to obtain an upper limit on the PEP violating parameter. 
The final result is that this parameter must be less than $5\times 
10^{-6}$. 

Other experiments are described in a review by  Gillaspy (2000). An 
incomplete list of additional references includes Barabash et al., 1998; 
Javorsek II et al., 2000, 2002; NEMO Collaboration (2000); Back et al. 
(Borexino), 2004; VIP Collaboration (2005). 

The validity of Bose statistics for photons has also been investigated 
(Ignatiev et al., 1996; DeMille et al., 1999).    

\section{Conclusion  and outlook}

To summarise, remarkable progress has been made since 1987 in the area of 
testing the Pauli exclusion principle and Bose statistics, both 
theoretically and experimentally. While essentially no measures of 
quantitative nature existed before 1987 on the accuracy of these 
principles, now the limits on possible statistics violation are quite 
impressive. For instance, the limit is $\sim 10^{-26}$ for electrons and 
even better for nucleons. Theoretically we have much better understanding 
of why PEP and Bose statistics are so accurate. We know what can go wrong 
when we try to violate them.  And we have important and interesting 
directions to pursue, both in theory and experiment. Finally, numerous 
links with other areas of physics and mathematics have been revealed. So, 
a lot of exciting work is ahead. 

I am grateful to V.A.Kuzmin for many helpful discussions of the subject 
of this paper. 


\begin{thebibliography}{99}
\bibitem{2} Amado, R.D., Primakoff, H., 1980.  Phys. Rev. C 22 (3) 1338-1340.
\bibitem{Borexino} Back, H.O. et al. (Borexino), 2004.  Eur. Phys. J. C 37,
  421-431. 
\bibitem{14} Barabash, A.S. et al., 1998. JETP Lett. 68, 112-116.
\bibitem{13} Biedenharn, L.C., Truini, P., van Dam, H., 1989.  J. Phys. A. Math.
  Gen. 22 (3), L67-71.
\bibitem{Biedenharn1989} Biedenharn, L.C., 1989.  J. Phys. A
  22 (18), L873.
\bibitem{Chow} Chow, C-K., Greenberg, O.W., 2001.  Phys. Lett. A 283 (1-2), 20-24.
\bibitem{72} Cougo-Pinto, M.V., 1993.  J. Math. Phys. 34
  (3), 1110-1124.
\bibitem{Cucurull:1995bx} Cucurull, L., Grifols, J.A., Toldr${\rm
    \grave{a}}$, R., 1996.  Astropart. Phys. 4 (4),
  391-395.
\bibitem{10} Deilamian, K., Gillaspy, J.D., Kelleher, D.E., 1995.
   Phys. Rev. Lett. 74 (24), 4787-4790.
\bibitem{DeMille} DeMille, D. et al., 1999. Phys. Rev. Lett., 83, 
3978-3981.
\bibitem{Dirac} Dirac, P.A.M., 1981.  In: Reidel, D. (Ed.), The impact of modern
  scientific ideas on society. Dordrecht, Holland, pp. 39-55.
\bibitem{Dolgov:2005qi} Dolgov, A.D., Smirnov, A.Y., 2005.  arXiv:hep-ph/0501066.
\bibitem{Dolgov:2005mi} Dolgov, A.D., Hansen, S.H., Smirnov, A.Y.,
  2005. 
  arXiv:astro-ph/0503612.
\bibitem{15} Drake, G.W.F, 1989.  Phys. Rev. A 39 (2), 897-899.
\bibitem{27} Fermi, E., 1933. Att. Sci. It. Progr. Sci. {\bf 22} Riunione
  (Bari ), vol. 3, p.7; Scientia {\bf 55}, 21, 1934.
\bibitem{Feynman} Feynman, R.P., Leighton, R.B., Sands, M., 1965.
The Feynman lectures on physics. Addison-Wesley, Reading. 
\bibitem{19} Fischbach, E., Kirsten, T., Shaeffer, O.Q., 1968.
  Phys.  Rev. Lett. 20 (18), 1012-1014.
\bibitem{100} Fivel, D.I., 1990.  Phys. Rev. Lett. 65,
  3361-3364; (E) 69, 2020, 1992.
\bibitem{8} Gavrin, V.N., Ignatiev, A.Yu., Kuzmin, V.A., 1988.  Phys. Lett. B 206,
  343-345.
\bibitem{Gillaspy} Gillaspy, J.D., 2000. In: Hilborn,
  R.C., Tino, G.M. (Eds.), AIP Conf. Proc. 545,
  241-252.
\bibitem{GillaspyHilborn}Gillaspy, J.D., Hilborn, R.C., 2000. 
http://physics.nist.gov/MajResFac/EBIT/peprefs.html.
\bibitem{117} Goldhaber, M., Scharff-Goldhaber, G., 1948.
  Phys. Rev. 73,  1472-1473.
\bibitem{26} Govorkov, A.B., 1983. Theor. Mat. Fiz. 54 (3), 361-371.
\bibitem{7} Govorkov, A.B., 1989.  Phys. Lett. A 137 (1-2), 7-10.
\bibitem{127} Green, H.S., 1953. Phys. Rev. 90 (2), 270-273.
\bibitem{131} Greenberg, O.W., Mohapatra, R.N., 1987. Phys.
  Rev. Lett. 59 (22), 2507-2510, (E) 61 (12), 1432 -1988.
\bibitem{134} Greenberg, O.W., Mohapatra, R.N., 1989. Phys. Rev. D 39 (7),
  2032-2038.
\bibitem{137} Greenberg, O.W., Mohapatra, R.N., 1989.  Phys. Rev. Lett. 62 (7), 712-714 (E) 62 (16), 1927 -1989.
\bibitem{138}Greenberg, O.W., 1990. Phys. Rev. Lett., 64, 705-708.
\bibitem{139} Greenberg, O.W., 1991. Phys. Rev. D 43 (12), 4111-4120.
\bibitem{Greenberg:2000zy} Greenberg, O.W., 2000. AIP Conf.\ Proc.\ 545, 113-127.
\bibitem{Hilborn}Hilborn, R.C., Tino, G.M., (Eds.), 2000. AIP Conf.\ Proc.\ 
545.
\bibitem{4} Ignatiev, A.Yu., Kuzmin, V.A., 1987. Yad. Fiz., 46, 786-790
  (Sov. J. Nucl. Phys. 47, 6, 1987).
\bibitem{4a} Ignatiev, A.Yu., Kuzmin, V.A., 1988. Pisma ZhETF, 47, 6-8.
   (JETP Lett., 47 (1), 4-7,1988).
\bibitem{17} Ignatiev, A.Yu., Kuzmin, V.A., 1989. In: Fackler, O., Tran
  Thanh Van, J. (Eds.), Tests of Fundamental Laws in Physics, Proc. IX
  Moriond Workshop, pp. 17-24.
\bibitem{Ignatiev} Ignatiev, A. Yu., 1990. Kyoto Preprint RIFP-854,
  http://ccdb3fs.kek.jp/cgi-bin/img\_index?9005309.
\bibitem{4b} Ignatiev, A.Yu., Joshi, J.C., Matsuda, M., 1996. Mod. Phys. Lett. A 11 (11), 871-876.
\bibitem{Javorsek2000} Javorsek II, D. et al., 2000. Phys. Rev. Lett. 85 (13), 2701-2704.
\bibitem{Javorsek2002} Javorsek II, D. et al., 2002. Nucl. Instr. and Meth. in
  Phys. Res. B 194 (1), 78-89.
\bibitem{Kaplan} Kaplan, I.G., 2002. Int. J. Quant.  Chemistry, 89 (4), 268-276.
\bibitem{Kinoshita} Kinoshita, J., 1988. Scientific American, June, p.19.
\bibitem{3} Kuzmin, V.A., 1984. In: Markov, M.A., Berezin, V.A., Frolov, V.P.
  (Eds.), Proc. of 3rd Seminar on Quantum Gravity, World Scientific,
  Singapore, pp. 270-288; NORDITA preprint 85/4, 1985.
\bibitem{20} Logan, B.A., Ljubicic, A., 1979. Phys. Rev. C 20 (5), 1957-1958.
\bibitem{28} Lyuboshitz, V.L., Podgoretskii, M.I. 1971. Sov. Phys. JETP 33 (1), 5-10.
\bibitem{Macfarlane1989} Macfarlane, A. J., 1989. J. Phys. A 22 (21), 4581-4588.
\bibitem{254} Mohapatra, R. N., 1990. Phys. Lett. B 242 (3-4), 407-411.
\bibitem{NEMO2000} NEMO Collaboration, 2000. Nucl. Phys. B (Proc. Suppl.) 87, 510.   
\bibitem{12a}Nolte, E. et al., 1991. Z. Phys. A, 340, 411.
\bibitem{12} Novikov, V.M., Pomansky, A.A., 1989. Pisma ZhETF, 49, 68.
\bibitem{21} Novikov, V.M. et al., 1990. Phys. Lett. B, 240, 227.
\bibitem{6} Okun, L.B., 1987. Pisma ZhETF 46 (11), 420-422, (JETP Lett., 46 (11),
  529-532, 1987).
\bibitem{29} Okun, L.B., 1988. In: Winter, K. (Ed.), Festival-Festschrift for
  Val Telegdi, ed. K. Winter, pp. 201-211.
\bibitem{18} Okun, L.B., 1989. Comments on Nucl. and Particle Phys. 19 (3), 99-116.
\bibitem{Pauli1955} Pauli, W., 1955. In: Pauli, W. (Ed.), Niels Bohr
  and the development of physics, London, Pergamon Press Ltd., pp.
  30-51.
\bibitem{22} Ramberg, E., Snow, G.A., 1990. Phys. Lett. B 238 (2-4), 438-441.
\bibitem{1a} Reines, F., Sobel, H.W., 1974. Phys. Rev. Lett. 32 (17), 954.
\bibitem{VIP} VIP Collaboration, 2005. 
http://www.lnf.infn.it/esperimenti/vip. 
\end{thebibliography}
\end{document}